\documentclass{PoS}
\title{Review of prospects for $H^+$ in non-SUSY multi Higgs models in view of ATLAS and CMS results}

\ShortTitle{Review of prospects for $H^+$ in non-SUSY multi Higgs models in view of ATLAS and CMS results}

\author{\speaker{Marc Sher}%
\\
        High Energy Theory Group\\
        Physics Department\\ College of William and Mary\\ Williamsburg, VA  23187 \\
        E-mail: \email{mtsher@wm.edu}}

\abstract{In this talk, prospects for the charged Higgs in non-SUSY models are reviewed, in view of LHC results (as of October, 2012).    The four models (Type I, Type II, lepton-specific and flipped) without tree level flavor-changing neutral currents are discussed.    Updates for the case in which the charged Higgs is lighter than the top, leading to production in top decays, are presented along with expectations for the future.    If the charged Higgs is heavier than the top quark, detection may be possible through the $\tau\nu$ decay mode.   In addition, it is pointed out that the decay of the charged Higgs into $W h$, where $h$ is the 125 GeV state, may be dominant.    Charged Higgs  phenomenology in models which do contain tree level flavor-changing neutral currents, as well as the inert doublet model and others, are also briefly reviewed.}

\FullConference{Prospects for Charged Higgs Discovery at Colliders\\
                 October 8-11, 2012\\
                 Uppsala University Sweden}

\begin{document}

\section{Introduction}

This workshop is the fourth in a series, and the first since data has begun coming in from ATLAS and CMS.   In this talk, I will review the status of the charged Higgs in various models, discuss current LHC results and future prospects, including the effects of a Standard Model-like neutral Higgs at 125 GeV.    Charged Higgs bosons in supersymmetric models will be discussed elsewhere - here the focus in on non-supersymmetric models.   The bulk of the talk will concentrate on models without flavor-changing neutral currents (FCNC), first looking at the case in which the charged Higgs is lighter than the top quark and then the case in which it is heavier.   I will then turn to models with FCNC, the inert model and others.     An very extensive review of two-Higgs-doublet models has appeared recently\cite{Branco:2011iw}, and I will use the notation of that review throughout.

In order to avoid tree-level FCNC, one can impose a $Z_2$ symmetry to ensure that all fermions of a given charge couple to a single Higgs boson\cite{Glashow:1976nt,Paschos:1976ay}.     This leads to four distinct models\cite{Barnett:1984zy,Barnett:1983mm,Barger:1989fj}.   In the Type I model, all fermions couple to one of the Higgs doublets; in the Type II model, the charge $2/3$ quarks couple to one doublet and the charge $-1/3$ quarks and leptons couple to the other.     There are two less studied versions.  In the lepton-specific model, the quarks couple to one doublet and the leptons to the other, and in the flipped model, the charge $2/3$ quarks and leptons couple to one doublet and the charge $-1/3$ quarks to the other.     These are summarized in Table 1.

\begin{table}[h]
\begin{center}
\begin{tabular}{|c|p{0.8in}|p{0.8in}|p{0.8in}|}  \hline
Model & $u_R^i$     & $d_R^i$& $e_R^i$     \\
\hline
Type I     & $\Phi_2$   &$ \Phi_2$ & $\Phi_2$      \\
\hline
Type II     & $\Phi_2$       &$\Phi_1$           & $\Phi_1$    \\
\hline
Lepton-specific     &$\Phi_2$    & $\Phi_2 $         & $\Phi_1$    \\
\hline
Flipped     & $\Phi_2$     & $\Phi_1$          & $\Phi_2$     \\
\hline
\end{tabular}
\end{center}
\caption{Models which lead to natural flavour conservation.
The superscript $i$ is a generation index.
By convention,
the $u^i_R$ always couples to $\Phi_2$.}
\end{table}

The Lagrangian describing the Yukawa interactions of the charged Higgs is
\begin{equation}
{\mathcal L}_{H^\pm} = - H^+ \left( \frac{\sqrt{2}\,V_{ud}}{v}\,
\bar{u} \left( m_u X P_L + m_d Y P_R \right) d +
\frac{\sqrt{2}\, m_\ell}{v}\,
Z \bar{\nu_L}\ell_R\right)
+ {\rm h.c.}
\end{equation}
The values of $X$, $Y$, and $Z$ depend on the particular model
and are given in Table 2.
\begin{table}[h]
\begin{center}
\begin{tabular}{|c|c|c|c|c|}  \hline
{} & Type I  & Type II & Lepton-specific & Flipped     \\
\hline
$X$ & $\cot{\beta}$ & $\cot{\beta}$ & $\cot{\beta}$ & $\cot{\beta}$ \\
\hline
$Y$ & $\cot{\beta}$ & $- \tan{\beta}$ & $\cot{\beta}$ & $- \tan{\beta}$ \\
\hline
$Z$ & $\cot{\beta}$ & $- \tan{\beta}$ & $- \tan{\beta}$ & $\cot{\beta}$ \\
\hline
\end{tabular}
\end{center}
\caption{Parameters describing the Yukawa coupling of the charged Higgs.}
\end{table}
Here, $\tan\beta$ is the ratio of the Higgs vacuum expectation values.   The phenomenology of the charged Higgs is thus very sensitive to the model (unless $\tan\beta=1$).

It should be noted that in the Type II and flipped models, radiative B decays force the charged Higgs to be heavier than $340$ GeV\cite{mahmoudi}.    But new physics can weaken this bound substantially (this happens in supersymmetric models, for example) and thus we will leave open the possibility that the charged Higgs could be much lighter.     We first consider the case in which the charged Higgs is lighter than the top quark.

\section{Charged Higgs lighter than the top quark}

If the charged Higgs is  lighter than the top quark, then $t\rightarrow bH^+$ can be a substantial decay mode.   Experimenters typically put bounds on the branching ratio for that process, and convert that into bounds in the $\tan\beta-M_{H^+}$ plane.    In making this conversion, they often specify a particular model, saying things like ``Interpreted in the context of the mmax h scenario of the MSSM".   While necessary for an extremely precise bound, the results of this conversion are, to leading order, much more general, applying to any Type II or flipped model.

\subsection{Type II model}

As can be seen from the Lagrangian, an upper bound on the Yukawa coupling in the Type II model leads to both a lower and an upper bound on $\tan\beta$, which varies with the charged Higgs mass.     The ATLAS results from the 2011 run of the LHC are shown in Figure 1.     The 2012 run will improve the reach, and possibly exclude masses in the 120-130 GeV range.   This figure presumes that the primary decay of the charged Higgs is into $\tau\nu$, which is the case for all $\tan\beta$ not very near 1.     Once the LHC is running at 13 TeV, and once hadronic tau decays are included, the entire range up to $150$ GeV will be excluded, if there is no signal (see the talk for several figures).    Alas, a $5\sigma$ discovery would be much more problematic, and may not be possible for many years at the LHC.

\vskip -1cm
\begin{figure}[htb]
\begin{center}
\includegraphics[width=0.50\textwidth]{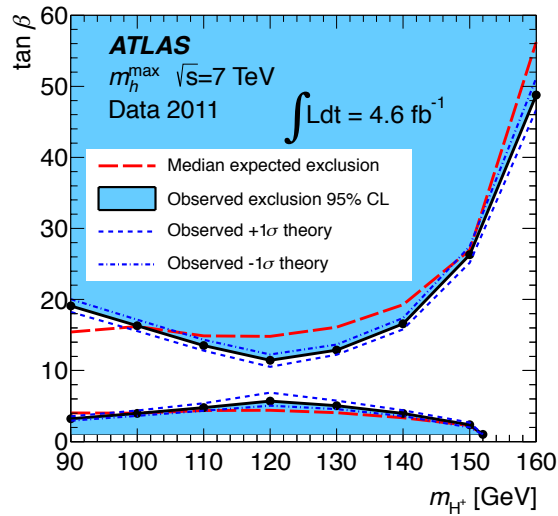}
\end{center}
\vskip -2cm
\caption{Bounds from $t\rightarrow bH^+$ in the Type-II model.}
\end{figure}

\subsection{Type I and lepton-specific models}
  
In the Type I and lepton-specific models, large values of $\tan\beta$ make the  charged Higgs fermiophobic or quarkphobic, respectively, and thus top decays will not be helpful.   The current results are very close to the lower line in Figure 1.   In the future, these bounds  will be substantially improved \cite{Aoki:2011wd,Guedes:2012eu}.    It was recently noted \cite{Mader:2012pm} that if the charged Higgs is very light, of the order of the $W$ mass, then associated production of a charged Higgs with a 125 GeV Higgs could be detected.

\subsection{Flipped model}

In the flipped model, the dominant decay of the charged Higgs \cite{Logan:2010ag} is into $bc$ and the subdominant decay is into cs.   Bounds from the Tevatron and LEP are fairly weak (excluding $\tan\beta$ values greater than 50 or less than 1.5 or so for much of the mass range).  An analysis for the LHC run has not yet been done, but one would expect the bounds to be weaker than in the other three models.

\section{Charged Higgs heavier than the top quark}

If the charged Higgs is heavier than the top quark, then the decay $H^+\rightarrow t\bar{b}$ becomes a dominant decay mode.    The recent discovery of a Higgs at $125$ GeV, however, opens up the possibility of $H^+\rightarrow h W^+$.    For the moment, let us ignore this possibility.    Observation of $H^+\rightarrow t\bar{b}$ is extremely difficult due to huge backgrounds, and thus most analyses have concentrated on the rarer, but cleaner, decay $H^+\rightarrow \tau\nu$.   The branching ratio for this decay, in the type II model, ranges from a few percent for moderate $\tan\beta$ to $10\%$ for large $\tan\beta$.

 \subsection{Fermionic decays}

In the Type II model, the main production mechanism is $gb\rightarrow tH^-$.     By looking at the $\tau\nu$ decay of the charged Higgs, one will be able to place bounds in the $\tan\beta-M_H$ plane.   For $30\ {\rm fb}^{-1}$, ATLAS projects excluding regions of $\tan\beta > 10$ for $M_H = 200$ GeV, rising to $\tan\beta > 50$ for $M_H= 500$ GeV  (see the talk for detailed figures).       The discovery reach is substantially weaker.    In the Type I and flipped models, the branching ratio into $\tau\nu$ is even smaller, making detection through this decay mode impossible.   The lepton-specific model has a huge branching ratio into $\tau\nu$, approaching $100\%$ for modest $\tan\beta$.   However, the quark induced production mechanisms are suppressed, and other production mechanisms are necessary.   The pair production process was been studied in Ref. \cite{Logan:2009uf} , and associated production in Ref.  \cite{Aoki:2009ha}.    The latter could be particularly relevant now that the $125$ GeV Higgs has been found---one could even look for $\mu^+\mu^-\tau\nu$ events, with the muon pair serving as a tag for the charged Higgs.     An alternative production mechanism was proposed by Eriksson et al.\cite{Eriksson:2006yt} and applies if the heavier scalar, $H$, or pseudoscalar, $A$, has a mass greater than that of the charged Higgs plus a W.   Then one can resonantly produce this scalar leading to a huge increase in associated production of an $H^+$ and a $W$.   While this does not generally occur in the MSSM, it certainly can occur in 2HDMs, and this possibility needs further analysis.

\subsection{$H^+\rightarrow hW^+$}

Perhaps the most interesting possible decay of a heavy charged Higgs is $H^+\rightarrow hW^+$.    The discovery of a Higgs at $125$ GeV makes this decay a viable option for charged Higgs bosons with masses over $210$ GeV.    The decay was first studied by Drees, et al.\cite{Drees:1999sb} and then by Assamagan, et al.\cite{Assamagan:2002ne}, however they only considered the hadronic decays of the light Higgs, finding potentially observable signals, but with substantial backgrounds.     A more recent analysis by Kanemura, et al.\cite{Kanemura:2009mk} looked at the branching ratios of the charged Higgs.     For a light Higgs of $120$ GeV, and a heavier neutral Higgs ($H$) mass of $150$ GeV, they found the result, in the Type II model for $\tan\beta = 3$, in Figure 2.    If the $H$ mass is heavier, similar in mass to the charged Higgs, then the upper line will vanish and one can see that the $hW$ branching ratio will exceed $t\bar{b}$ for masses above $350$ GeV, and will be substantial for lower masses.

\begin{figure}[htb]
\begin{center}
\includegraphics[width=0.8\textwidth]{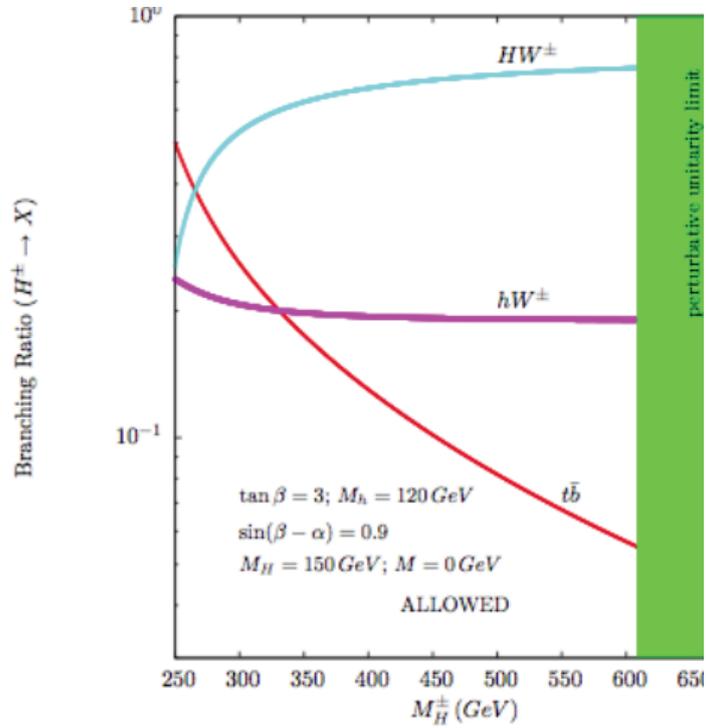}
\end{center}
\vskip -3cm
\caption{Branching ratios of the charged Higgs in the Type II model.}
\end{figure}

We see that the $H^+\rightarrow hW^+$ decay may be dominant in the Type II model.   In the Type I model, the charged Higgs is fermiophobic for moderately large $\tan\beta$, and thus the $t\bar{b}$ is suppressed, leading to a much higher branching ratio for $H^+\rightarrow hW^+$.  However, the $gb\rightarrow tH^-$ production mechanism is suppressed as well and thus other production mechanisms are necessary.
In the lepton-specific model, the domination of the $\tau\nu$ decay of the charged Higgs keeps the $hW$ branching fraction at below $1\%$, and in the flipped model the branching fraction is even smaller.      A much more detailed analysis of this decay is needed.

\section{Other models}

If one does not insist on avoiding FCNC, the most general Yukawa couplings  in a two-Higgs doublet model are (listing quarks only, for simplicity)
\begin{equation}
\bar{Q}_L\eta_1^UU_R\tilde{\Phi}_1+\bar{Q}_L\eta_1^DD_R{\Phi}_1+\bar{Q}_L\eta_2^UU_R\tilde{\Phi}_2+\bar{Q}_L\eta_2^DD_R{\Phi}_2
\end{equation}
where the $\eta$'s are Yukawa coupling matrices in flavor-space.    Following the notation of Mahmoudi and Stal \cite{Mahmoudi:2009zx}, we define
\begin{equation}
\kappa^F \equiv \eta_1^F \cos\beta + \eta_2^F \sin\beta
\end{equation}
\begin{equation}
\rho^F \equiv -\eta_1^F\sin\beta +\eta_2^F\cos\beta
\end{equation}
Here, one Higgs has a vacuum expectation value and diagonal Yukawa couplings $\kappa^F$ and the other has no vacuum expectation value and Yukawa couplings  $\rho^F$, which might be flavor-changing.

The Cheng-Sher ansatz \cite{Cheng:1987rs} was proposed to suppress (but not eliminate) FCNC in a natural way:
\begin{equation}
\rho^F_{ij} = \lambda_{ij}^F \frac{\sqrt{2m_im_j}}{v}
\end{equation}
where the $\lambda_{ij}$ are $O(1)$ (although some argue that $O(1/\tan\beta)$ is more appropriate).    After 25 years, this model is under serious challenge -- current bounds (using a pseudoscalar mass of $400$ GeV) from meson-antimeson mixing are \cite{Golowich:2007ka}
\begin{equation}
(\lambda_{ds},\lambda_{uc},\lambda_{bd},\lambda_{bs}) \leq (0.3, 0.6, 0.18, 0.18)
\end{equation}
The dominant charged Higgs vertex will be into $c_Rb_L$, unlike $c_Lb_R$ in the Type I and II models, and is larger, of $O(1)\%$.   This leads \cite{He:1998ie} to the possibility of s-channel charged Higgs production, substantially increasing the rate of single top production and $hW$ production.   For masses below $350$ GeV, this should be observable.  However, this work is 15 years old, and is in need of updating.

The aligned two-Higgs doublet model \cite{Pich:2009sp} simply assumes that the $\rho^F$ and $\eta^F$ matrices are proportional, leading to no tree-level FCNC.  This assumption is not radiatively stable, but provides a convenient parametrization with the four models above as special cases.   See Ref. \cite{Mahmoudi:2009zx} for a discussion and limits from low-energy processes.

The inert doublet model eliminates all of the couplings of a Higgs doublet to fermions and also does not allow a vacuum value for this doubleet.  The lightest star is then a dark matter candidate.   The charged Higgs will only decay via $HW$ and $AW$ (the lightest of H or A is then stable).   Miao, et al. \cite{Miao:2010rg} studied these decays and concluded that charged Higgs masses above 120 GeV could not be detected due to large backgrounds.

Finally, the neutrino-specific model \cite{Davidson:2009ha} has one Higgs coupling only to neutrinos, giving them a Dirac mass.  The vacuum value of the Higgs is O(eV).  The only decay mode of the charged Higgs is then into a lepton and a neutrino.   For a normal hierarchy, it is primarily into $\tau\nu$ and $\mu\nu$, whereas for an inverted hierarch into $e\nu$.  The signature for $pp\rightarrow H^+H^- \rightarrow \ell^+\ell^-\nu\nu$ is quite dramatic and a study of various cuts shows a $5\sigma$ discovery range up to 300 GeV for $100/{\rm fb}$ at $14$ TeV.

\section{Conclusions}

Studies of the phenomenology of charged Higgs bosons tend to fall into two categories:   those in which the charged Higgs is produced in top quark decays, and those in which the charged Higgs is heavier than the top.   A nice feature of charged Higgs phenomenology is that the couplings to fermions only depend on a single parameter, $\tan\beta$, with the dependence being sensitive to the specific two-Higgs doublet model.

If the charged Higgs is produced in top quark decays, then the current run will be sensitive to substantial regions of parameter-space.   In the Type II model, a mass range around 135 GeV can be excluded (unless there is a discovery, of course) for all $\tan\beta$, and shortly after the 2014 start-up, the entire range up to $150$ GeV can be excluded.   In the Type I and lepton-specific models, only lower bounds on $\tan\beta$ can be obtained; for larger $\tan\beta$ discovery may not be possible at the LHC.    In the flipped model, the dominant decay of the charged Higgs is into $c\bar{b}$, and more studies are needed to get significant constraints.

If the charged Higgs is heavier than the top quark,  then in the Type II model one can obtain an upper bound on $\tan\beta$ by exploring the $\tau\nu$ decay of the charged Higgs.   However, the most promising decay might be $h W$, for which detailed analyses with a $125$ GeV Higgs have yet to be done.   The Type I model is similar, but one needs non-fermionic production for large $\tan\beta$;  more studies are needed.   In the lepton-specific case, one might get multi-tau events.  There does not appear to be any hope of detection of a charged Higgs in the flipped model.

In models with tree-level flavor-changing neutral currents, one might get a substantial production rate due to s-channel production of a charged Higgs.   This might be observable at the LHC upgrade.   Other models were also briefly discussed.

I would like to thank the organizers for their invitation.   This work was supported by a Joseph Plumeri Award and by the National Science Foundation under grant PHY-1068008.


\begin{thebibliography}{99}
\bibitem{Branco:2011iw}
  G.~C.~Branco, P.~M.~Ferreira, L.~Lavoura, M.~N.~Rebelo, M.~Sher and J.~P.~Silva,
  Phys.\ Rept.\  {\bf 516} (2012) 1
  [arXiv:1106.0034 [hep-ph]].
  \bibitem{Glashow:1976nt}
 S.~L.~Glashow and S.~Weinberg,
 Phys.\ Rev.\  D {\bf 15} (1977) 1958.
\bibitem{Paschos:1976ay}
 E.~A.~Paschos,
 Phys.\ Rev.\  D {\bf 15} (1977) 1966.
\bibitem{Barnett:1984zy}
  R.~M.~Barnett, G.~Senjanovic and D.~Wyler,
  Phys.\ Rev.\ D {\bf 30} (1984) 1529.
\bibitem{Barnett:1983mm}
  R.~M.~Barnett, G.~Senjanovic, L.~Wolfenstein and D.~Wyler,
  Phys.\ Lett.\ B {\bf 136} (1984) 191.
\bibitem{Barger:1989fj}
  V.~D.~Barger, J.~L.~Hewett and R.~J.~N.~Phillips,
  Phys.\ Rev.\ D {\bf 41} (1990) 3421.
  \bibitem{mahmoudi}
  F. Mahmoudi, talk given at Prospects For Charged Higgs Discovery At Colliders (CHARGED 2012), 8-11 October, Uppsala, Sweden.
\bibitem{Aoki:2011wd}
  M.~Aoki, R.~Guedes, S.~Kanemura, S.~Moretti, R.~Santos and K.~Yagyu,
  Phys.\ Rev.\ D {\bf 84} (2011) 055028
  [arXiv:1104.3178 [hep-ph]].
\bibitem{Guedes:2012eu}
  R.~Guedes, S.~Moretti and R.~Santos,
  JHEP {\bf 1210} (2012) 119
  [arXiv:1207.4071 [hep-ph]].
  \bibitem{Mader:2012pm}
  W.~Mader, J.~-h.~Park, G.~M.~Pruna, D.~Stockinger and A.~Straessner,
  JHEP {\bf 1209} (2012) 125
  [arXiv:1205.2692 [hep-ph]].
\bibitem{Logan:2010ag}
  H.~E.~Logan and D.~MacLennan,
  Phys.\ Rev.\ D {\bf 81} (2010) 075016
  [arXiv:1002.4916 [hep-ph]].
\bibitem{Logan:2009uf}
  H.~E.~Logan and D.~MacLennan,
  Phys.\ Rev.\ D {\bf 79} (2009) 115022
  [arXiv:0903.2246 [hep-ph]].
\bibitem{Aoki:2009ha}
  M.~Aoki, S.~Kanemura, K.~Tsumura and K.~Yagyu,
  Phys.\ Rev.\ D {\bf 80} (2009) 015017
  [arXiv:0902.4665 [hep-ph]].
\bibitem{Eriksson:2006yt}
  D.~Eriksson, S.~Hesselbach and J.~Rathsman,
  Eur.\ Phys.\ J.\ C {\bf 53} (2008) 267
  [hep-ph/0612198].
\bibitem{Drees:1999sb}
  M.~Drees, M.~Guchait and D.~P.~Roy,
  Phys.\ Lett.\ B {\bf 471} (1999) 39
  [hep-ph/9909266].
\bibitem{Assamagan:2002ne}
  K.~A.~Assamagan, Y.~Coadou and A.~Deandrea,
  Eur.\ Phys.\ J.\ direct C {\bf 4} (2002) 9
  [hep-ph/0203121].
\bibitem{Kanemura:2009mk}
  S.~Kanemura, S.~Moretti, Y.~Mukai, R.~Santos and K.~Yagyu,
  Phys.\ Rev.\ D {\bf 79} (2009) 055017
  [arXiv:0901.0204 [hep-ph]].
\bibitem{Mahmoudi:2009zx}
  F.~Mahmoudi and O.~Stal,
  Phys.\ Rev.\ D {\bf 81} (2010) 035016
  [arXiv:0907.1791 [hep-ph]].
\bibitem{Cheng:1987rs}
  T.~P.~Cheng and M.~Sher,
  Phys.\ Rev.\ D {\bf 35} (1987) 3484.
\bibitem{Golowich:2007ka}
  E.~Golowich, J.~Hewett, S.~Pakvasa and A.~A.~Petrov,
  Phys.\ Rev.\ D {\bf 76} (2007) 095009
  [arXiv:0705.3650 [hep-ph]].
\bibitem{He:1998ie}
  H.~-J.~He and C.~P.~Yuan,
  Phys.\ Rev.\ Lett.\  {\bf 83} (1999) 28
  [hep-ph/9810367].
\bibitem{Pich:2009sp}
  A.~Pich and P.~Tuzon,
  Phys.\ Rev.\ D {\bf 80} (2009) 091702
  [arXiv:0908.1554 [hep-ph]].
\bibitem{Miao:2010rg}
  X.~Miao, S.~Su and B.~Thomas,
  Phys.\ Rev.\ D {\bf 82} (2010) 035009
  [arXiv:1005.0090 [hep-ph]].
\bibitem{Davidson:2009ha}
  S.~M.~Davidson and H.~E.~Logan,
  Phys.\ Rev.\ D {\bf 80} (2009) 095008
  [arXiv:0906.3335 [hep-ph]].




\end{thebibliography}
\end{document}